\documentclass[conference,letterpaper]{IEEEtran}
\usepackage{epsfig,amsfonts,subfigure,color,amsbsy}
\usepackage[nolist]{acronym}
\usepackage{graphicx,cite,amssymb,amsthm}
\usepackage{amsmath}
\usepackage{mathrsfs}
\usepackage{tikz}
\usepackage{multirow}
\usepackage{bigstrut}
\usepackage{psfrag}
\usepackage{stackengine}
\usetikzlibrary{shapes,arrows}
\usepackage{xcolor}
\usepackage{mathtools}
\usepackage{tabularx,array}
\usepackage{mathbbol}
\usepackage{lipsum}
\mathtoolsset{showonlyrefs}
\DeclareMathOperator*{\argmin}{argmin}
\DeclareMathOperator*{\argmax}{argmax}

\IEEEoverridecommandlockouts

\begin{document}
\newtheorem{thm}{Theorem}
\newtheorem{lemma}{Lemma}
\newtheorem{definition}{Definition}
\newtheorem{example}{Example}
\newtheorem{remark}{Remark}

\newcommand{\expect}[1]{\mathbb{E}[#1]}
\newcommand{\expectation}{\mathbb{E}}
\newcommand{\figwidth}{1\columnwidth}
\newcommand{\figwidthsmall}{0.7\columnwidth}
\newcommand{\todo}[1]{} 
\newcommand{\msr}[1]{\mathscr{#1}}
\newcommand{\KL}{\mathcal{D}}
\newcommand{\PCM}{\mathbf{H}}
\newcommand{\field}{\mathbb{F}}
\newcommand{\PCE}{\mathcal{C}}
\newcommand{\BEF}{H_b}
\newcommand{\cu}{\stackrel{\!\!\raisebox{-2pt}{\tiny{$\mathrm{u}$}}}{\rightarrow}}
\newcommand{\f}{\mathsf{f}}
\newcommand{\g}{\mathsf{g}}
\newcommand{\h}{\mathsf{h}}
\newcommand{\MLthr}{\epsilon_{\mathsf{ML}}^{*}}
\newcommand{\ri}{r_{\mathsf{i}}}
\newcommand{\ro}{r_{\mathsf{o}}}
\newcommand{\cP}{\mathcal{P}}

\definecolor{gl}{rgb}{0.0,0.5,0.8}
\definecolor{fc}{rgb}{0.8,0.5,0}
\definecolor{al}{rgb}{1,0.3,0.3}
\newcommand{\giangio}{\textcolor{gl}}
\newcommand{\fede}{\textcolor{fc}}
\newcommand{\alert}{\textcolor{al}}



 
\title{A Lower Bound on the Error Exponent\\ of Linear Block Codes over the Erasure Channel}

\author{%
  \IEEEauthorblockN{Enrico Paolini}
  \IEEEauthorblockA{DEI, University of Bologna\\
                    via dell'Universit{\`a} 50, Cesena (FC) Italy\\
                    Email: e.paolini@unibo.it}
  \and
  \IEEEauthorblockN{Gianluigi Liva}
  \IEEEauthorblockA{KN-SAN, German Aerospace Center\\
                    M{\"u}nchener Strasse 20, Wessling, Germany\\
                    Email: Gianluigi.Liva@dlr.de}
}

 \maketitle

\thispagestyle{empty}


\begin{abstract}
A lower bound on the \ac{ML} decoding error exponent of linear block code ensembles, on the erasure channel, is developed. The lower bound turns to be positive, over an ensemble specific interval of erasure probabilities, when the ensemble weight spectral shape function tends to a negative value as the fractional codeword weight tends to zero. For these ensembles we can therefore lower bound the block-wise ML decoding threshold. Two examples are presented, namely, linear random parity-check codes and fixed-rate Raptor codes with linear random precoders. While for the former a full analytical solution is possible, for the latter we can lower bound the \ac{ML} decoding threshold on the erasure channel by simply solving a \boldmath{$2 \times 2$} system of nonlinear equations.
\end{abstract}

\thispagestyle{empty}
\setcounter{page}{1}




\begin{acronym}
\acro{$q$-EC}{$q$-ary erasure channel}
\acro{BEC}{binary erasure channel}
\acro{BP}{belief propagation}
\acro{CRDSA}{contention resolution diversity slotted Aloha}
\acro{DE}{density evolution}
\acro{EXIT}{extrinsic information transfer}
\acro{i.i.d.}{independent and identically distributed}
\acro{IC}{interference cancellation}
\acro{IRSA}{irregular repetition slotted Aloha}
\acro{KL}{Kullback-Leibler}
\acro{LDPC}{low-density parity-check}
\acro{LHS}{left-hand side}
\acro{LT}{Luby-transform}
\acro{MAC}{medium access control}
\acro{MAP}{maximum a posteriori probability}
\acro{PLR}{packet loss rate}
\acro{r.v.}{random variable}
\acro{RHS}{right-hand side}
\acro{SINR}{signal-to-interference-and-noise ratio}
\acro{SNR}{signal-to-noise ratio}
\acro{p.m.f.}{probability mass function} 
\acro{SIC}{successive interference cancellation} 
\acro{$q$-EC}{$q$-ary erasure channel}
\acro{ML}{maximum likelihood}
\acro{WEF}{weight enumerating function}
\acro{MDS}{maximum distance separable}
\acro{AWE}{average weight enumerator}
\acro{DMC}{discrete memory-less channel}
\acro{EC}{erasure channel}
\acro{w.r.t.}{with respect to}
\end{acronym}


\section{Introduction}\label{sec:not}

In this paper\footnote{A shorter version of this paper, omitting some proofs, has been submitted to the 2019 IEEE International Symposium on Information Theory (ISIT).}, a lower bound on the \ac{ML} decoding error exponent of linear code ensembles when used over \acp{EC} is derived. The calculation of the bound requires the knowledge of the ensemble weight spectral shape only (under a relatively mild condition, as it will be discussed later).
A general lower bound on the error exponent, for any \ac{DMC}, was introduced \cite{Shulman99:random}. Its calculation involves the evaluation of the maximum ratio between the ensemble \ac{AWE} and the \ac{AWE} of the random linear code ensemble. The technique of \cite{Shulman99:random} was used in \cite{Miller01:Bounds_ML_LDPC} to derive a lower bound on the \ac{ML} decoding error exponent of (expurgated) \ac{LDPC} code ensembles \cite{studio3:GallagerBook}. 

 The bound on the error exponent introduced  in this paper is  derived from the tight union bound on the error probability under \ac{ML} decoding over the \ac{EC} for linear block code ensembles of \cite{C.Di2001:Finite,Liva13:bounds}. A similar approach was followed in \cite{Burshten04:Asymptotic} to obtain a lower bound on the error exponent for expurgated \ac{LDPC} code ensembles. Our work extends the result of \cite{Burshten04:Asymptotic} to any linear code ensemble for which the weight spectral shape is known, with the only requirement that the logarithm of the \ac{AWE} of the code ensemble (normalized to the block length) converges in the block length uniformly  to the weight spectral shape. The lower bound turns out to be positive, over an ensemble specific interval of erasure probabilities, when the ensemble weight spectral shape function tends to a negative value as the fractional codeword weight tends to zero. For the linear random code ensemble, we show that the bound on the error exponent recovers Gallager's random coding error exponent \cite{studio3:GallagerBound}.
The knowledge of the lower bound on the error exponent allows obtaining a lower bound on the ensemble's \ac{ML} erasure decoding threshold. As an example of application, we derive a lower bound on \ac{ML} erasure decoding threshold for the ensemble of fixed-rate Raptor codes \cite{shokrollahi06:raptor} introduced in \cite{Lazaro16:Raptor}. Remarkably, the result is obtained by simply solving a $2 \times 2$ system of nonlinear equations. For the analyzed ensembles, the bound on the error exponent derived in this paper shows to be considerably tighter than the general bound of \cite{Shulman99:random} when the latter is specialized to the \ac{BEC}.

\section{Preliminaries}\label{sec:prel}

We consider transmission of linear block codes constructed over $\mathbb F_q$, the finite field of order $q$, on a memoryless \ac{$q$-EC} on which each codeword symbol is correctly received with probability $1-\epsilon$ and erased with probability $\epsilon$. A code ensemble is defined as a set of codes along with a probability distribution on such codes. We denote by $\mathcal{C}(n,r,q)$ a generic ensemble of linear block codes over $\mathbb F_q$ of length $n$ and design rate $r$, and by $\mathsf C \in \mathcal{C}(n,r,q)$ a random code in the ensemble. The block-wise \ac{ML} decoding error probability of $\mathsf{C}$ over the \ac{$q$-EC} is indicated as $P_B(\mathsf{C},\epsilon)$ and its expectation over the ensemble as $\expectation_{\mathcal{C}(n,r,q)}[P_B(\mathsf{C},\epsilon)]$. We define the \ac{ML} decoding threshold for the ensemble $\mathcal{C}(n,r,q)$ over the \ac{$q$-EC} as $\MLthr=\sup\{\epsilon \in (0,1) : \expectation_{\mathcal{C}(n,r,q)}[P_B(\mathsf{C},\epsilon)] \rightarrow 0 \textrm{ as } n \rightarrow \infty \}$.

Our starting point is an upper bound on $\expectation_{\mathcal{C}(n,r,q)}[P_B(\mathsf{C},\epsilon)]$ developed in \cite{C.Di2001:Finite} for binary codes and extended in \cite{Liva13:bounds} to non-binary ones. We have
\begin{align}\label{eq:Di_q}
&\!\!\!\!\!\!\!\!\expectation_{\mathcal{C}(n,r,q)} \left[P_B(\mathsf{C},\epsilon)\right]\leq \sum_{e=(1-r)n+1}^{n} {n \choose e}
\epsilon^e (1-\epsilon)^{n-e} \notag \\ 
&\!\!\!\!\!\!\!\!\!\!+ \sum_{e=1}^{(1-r)n}\!\! {n \choose e}
\epsilon^e (1-\epsilon)^{n-e} \min \Bigg\{1, \frac{1}{q-1}\sum_{w=1}^e {e \choose w}
\frac{\mathcal A _w}{{n \choose w}}\Bigg\}
\end{align}
where $\mathcal A(x) = \sum_{i=0}^n \mathcal A_i x^i$ is the \ac{AWE} of $\mathsf C$. Given the \ac{AWE} $\mathcal A(x)$ the growth rate of the weight distribution, or weight spectral shape, of $\mathcal{C}(n,r,q)$ is defined as\footnote{In this paper all logarithms are to the base $2$.} $G(\omega)=\lim_{n \rightarrow \infty} \frac{1}{n} \log \mathcal{A}_{\lfloor \omega n \rfloor}$.

We denote the \ac{KL} divergence between two Bernoulli distributions with parameters $u$ and $v$, both in $(0,1)$, by $\KL(u,v)=u \log \frac{u}{v}+(1-u)\log \frac{1-u}{1-v}$. %
%
%
Moreover, we denote by $\BEF(u)=-u \log u - (1-u) \log (1-u)$, $0 \leq u \leq 1$, the binary entropy function. Throughout the paper we make use of the lower and upper bounds
\begin{align}\label{eq:bin_coeff_bounds}
\frac{1}{n+1} 2^{n\BEF(k/n)} \leq {n \choose k} \leq 2^{n\BEF(k/n)}
\end{align}
on the binomial coefficient, valid for all nonnegative integers $k \leq n$. For any two pairs $(x_1,y_1)$ and $(x_2,y_2)$ of reals, we write $(x_1,y_1) \succeq (x_2,y_2)$ when $x_1 \geq x_2$ and $y_1 \geq y_2$.

Recall that a sequence $f_n$ of real-valued functions on $A \subseteq \mathbb R$ converges uniformly to the function $f : A \mapsto \mathbb R$ on $A_0 \subseteq A$ if for any $\varepsilon > 0$ there exists $n_0(\varepsilon)$ such that, for all $n \geq n_0(\varepsilon)$, $| f_n(x) - f(x) | < \varepsilon$ $\forall x \in A_0$.
We write $f_n \cu f$ to indicate that $f_n$ converges to $f$ uniformly. A necessary and sufficient condition for uniform convergence is established by the following lemma \cite[Th.~7.10]{RudinBook}.
\begin{lemma}\label{lemma:supmetric}
Let $\lim_{n \rightarrow \infty} f_n(x) = f(x)$ $\forall x \in A_0$. Then $f_n \cu f$ on $A_0$ if and only if $\sup_{x \in A_0} | f_n(x) - f(x) | \rightarrow 0$ as $n \rightarrow \infty$.
\end{lemma}
The following result will also be useful.
\begin{lemma}\label{lemma:absinfbound}
Let $f,g: A \subseteq \mathbb R \mapsto \mathbb R$ be bounded functions. Then
\begin{align*}
\Big| \inf_{x \in A} f(x) - \inf_{x \in A} g(x) \Big| \leq \sup_{x \in A} \left| f(x) - g(x) \right| \, .
\end{align*}
\end{lemma}

\section{Main Results}\label{sec:exp}

This section presents the main results of this paper. A lower bound on the asymptotic error exponent on linear block code ensembles over the erasure channel is first developed in Theorem~\ref{theorem:lower_bound}. Then, Theorem~\ref{thm:delta_star} shows how this bound allows lower bounding $\MLthr$ for ensembles for which $G(\omega)$ is continuous in $(0,1]$ and negative for small enough $\omega$.

\begin{thm}\label{theorem:lower_bound}
Consider a linear block code ensemble $\mathcal{C}(n,r,q)$ and let its weight spectral shape $G(\omega)$ be well-defined in $[0,1]$. If $\frac{1}{n} \log \mathcal A_{\lfloor \omega n \rfloor} \cu G(\omega)$ then
\begin{align*}
\lim_{n \rightarrow \infty} - \frac{1}{n} \log \expectation_{\mathcal{C}(n,r,q)} \left[P_B(\mathsf{C},\epsilon)\right] \geq E_{G}(\epsilon)
\end{align*}
where
\begin{align}\label{eq:exp_lower_bound}
E_{G}(\epsilon) =  \inf_{\delta \in (0,1]} \f_{\epsilon}(\delta) \, .
\end{align}
The function $\f_{\epsilon}(\delta)$ is defined as
\begin{align}\label{eq:feps}
\f_{\epsilon}(\delta) &= \KL(\delta,\epsilon) + \g^+(\delta) 
\end{align}
where 
\begin{align}\label{eq:gplus}
\g^+(\delta)=\max \{0, \g(\delta)\}
\end{align}
and
\begin{align}\label{eq:g}
\g(\delta) &= \inf_{\omega \in (0,\delta]} \Big[ -\delta \BEF\Big(\frac{\omega}{\delta}\Big)+\BEF(\omega) - G(\omega) \Big] \, .
\end{align}
\end{thm}
\begin{IEEEproof}
The proof is organized into two parts. We first upper bound the right-hand side of \eqref{eq:Di_q} to obtain a lower bound on $-\frac{1}{n} \log \expectation_{\mathcal{C}(n,r,q)} \left[P_B(\mathsf{C},\epsilon)\right]$. Then we take the limit of the lower bound as $n \rightarrow \infty$.

\subsubsection{Lower bounding $-\frac{1}{n} \log \expectation_{\mathcal{C}(n,r,q)} \left[P_B(\mathsf{C},\epsilon)\right]$}
The upper bound \eqref{eq:Di_q} can be written in the equivalent, more compact form
\begin{align}\label{eq:Di_q_compact}
& \expectation_{\mathcal{C}(n,r,q)} \left[P_B(\mathsf{C},\epsilon)\right] \notag \\
&\leq \sum_{e=1}^{n} {n \choose e}
\epsilon^e (1-\epsilon)^{n-e} \min \Bigg\{1, \frac{1}{q-1}\sum_{w=1}^e {e \choose w}
\frac{\mathcal A _w}{{n \choose w}}\Bigg\} .
\end{align}
Letting $w=\omega n$ and $e=\delta n$, we have\footnote{For notational simplicity hereafter we write $\mathcal{A}_{\omega n}$ in lieu of~$\mathcal{A}_{\lfloor \omega n \rfloor}$.}
\begingroup
\allowdisplaybreaks
\begin{align}\label{eq:UB_E_Pb}
&\expectation_{\PCE(n,r,q)} \left[P_B(\mathsf{C},\epsilon)\right] \notag \\ 
& \stackrel{\mathrm a}{\leq} n \max_{e \in \mathbb \{1,\dots,n\}} \bigg[ {n \choose e} \epsilon^{e} (1-\epsilon)^{n-e}  \notag \\ 
& \times \min \bigg\{1, \frac{1}{q-1} \sum_{w=1}^e {e \choose w}
\frac{\mathcal A _w}{{n \choose w}} \bigg\} \bigg] \notag \\ 
& \stackrel{\mathrm b}{\leq} n \max_{e \in \mathbb \{1,\dots,n\}} \bigg[ {n \choose e} \epsilon^{e} (1-\epsilon)^{n-e}  \notag \\ 
& \times \min \bigg\{1, \frac{e}{q-1} \max_{w \in \mathbb \{1,\dots,e\}} \bigg( {e \choose w}
\frac{\mathcal A _w}{{n \choose w}} \bigg) \bigg\} \bigg] \notag \\ 
&= n \max_{\delta \in \{\frac{1}{n}, \dots, 1\}} \bigg[ {n \choose \delta n} \epsilon^{\delta n} (1-\epsilon)^{n(1-\delta)} \notag \\
&\times \min \bigg\{1, \frac{\delta n}{q-1} \max_{\omega \in \{\frac{1}{n},\dots,\delta\}} \bigg( {\delta n \choose \omega n} \frac{\mathcal A _{\omega n}}{{n \choose \omega n}} \bigg) \bigg\} \bigg] \notag \\
& \stackrel{\mathrm c}{\leq} n \max_{\delta \in \{\frac{1}{n}, \dots, 1\}} \bigg[ 2^{n (\BEF(\delta)+ \delta \log \epsilon+(1-\delta)\log(1-\epsilon))} \notag \\
& \times \min \bigg\{ 1, \frac{\delta n(n+1)}{q-1}\!\! \max_{\omega \in \{\frac{1}{n},\dots,\delta\}} \!\! 2^{n( \delta \BEF(\frac{\omega}{\delta}) - \BEF(\omega) + \frac{\log \mathcal A _{\omega n}}{n})} \bigg\}\!\bigg] \notag \\
& \stackrel{\mathrm d}{=} n \max_{\delta \in \{\frac{1}{n}, \dots, 1\}} \bigg[ 2^{-n \KL(\delta,\epsilon)}  \notag \\
& \times \min \bigg\{ 1, \frac{\delta n(n+1)}{q-1}\!\! \max_{\omega \in \{\frac{1}{n},\dots,\delta\}} \!\! 2^{n( \delta \BEF(\frac{\omega}{\delta}) - \BEF(\omega) + \frac{\log \mathcal A _{\omega n}}{n})} \bigg\}\!\bigg] \notag \\
& \stackrel{\mathrm e}{\leq} n \sup_{\delta \in \mathbb Q \cap (0,1]} \bigg[ 2^{-n \KL(\delta,\epsilon)}  \notag \\
& \times \min \bigg\{ 1, \frac{\delta n(n+1)}{q-1}\!\! \sup_{\omega \in \mathbb Q \cap (0,\delta]} \!\! 2^{n( \delta \BEF(\frac{\omega}{\delta}) - \BEF(\omega) + \frac{\log \mathcal A _{\omega n}}{n})} \bigg\}\!\bigg] \notag \\
& \stackrel{\mathrm f}{=} n \sup_{\delta \in (0,1]} \bigg[ 2^{-n \KL(\delta,\epsilon)}  \notag \\
& \times \min \bigg\{ 1, \frac{\delta n(n+1)}{q-1}\!\! \sup_{\omega \in (0,\delta]} \!\! 2^{n( \delta \BEF(\frac{\omega}{\delta}) - \BEF(\omega) + \frac{\log \mathcal A _{\omega n}}{n})} \bigg\}\!\bigg]
\end{align}
\endgroup
In the above development: `$\mathrm a$' and `$\mathrm b$' are due to $\sum_{l=1}^h f(l) \leq h \max_{l\in \mathbb N^*_{h}} f(l)$. Moreover:  `$\mathrm c$' is due to application of the upper and lower bounds in \eqref{eq:bin_coeff_bounds}; `$\mathrm d$' to  expanding $\BEF(\delta)$ and recalling the definition of \ac{KL} divergence; `$\mathrm e$' to the fact that the supremum over $\mathbb Q \cap (0,1]$ upper bounds the maximum over $\{\frac{1}{n},\dots,1\}$ and, similarly, the supremum over $\mathbb Q \cap (0,\delta]$ upper bounds the maximum over $\{\frac{1}{n},\dots,\delta\}$; `$\mathrm f$' to the density of $\mathbb Q$. In the final expression, both $\delta$ and $\omega$ are considered as real variables. The bound \eqref{eq:UB_E_Pb} is valid for any length $n$, rate $r$, and field order $q$. 

Next we exploit \eqref{eq:UB_E_Pb} to bound $-\frac{1}{n} \log \expectation_{\mathcal{C}(n,r,q)} \left[P_B(\mathsf{C},\epsilon)\right]$ from below. Owing to logarithm monotonicity we obtain
\begin{align}\label{eq:bound_log}
&-\frac{1}{n} \log \expectation_{\mathcal{C}(n,r,q)} \left[P_B(\mathsf{C},\epsilon)\right] \geq \inf_{\delta\in(0,1]} \f_n(\delta) .
\end{align}
where
\begin{align}\label{eq:fn}
\f_{\epsilon,n}(\delta) &= -\frac{1}{n} \log n +\KL(\delta,\epsilon) + \max \bigg\{ 0, \inf_{\omega\in(0,\delta]} \bigg( \frac{\log\frac{q-1}{\delta n(n+1)}}{n} \notag \\
&\phantom{....} -\delta \BEF\left(\frac{\omega}{\delta}\right) + \BEF(\omega) - \frac{\log \mathcal A _{\omega n}}{n}  \bigg) \bigg\}.
\end{align}

\subsubsection{Taking the limit}

Next, we take the limit as $n \rightarrow \infty$ in both sides of \eqref{eq:bound_log}. To keep the notation compact we define
\begin{align*}
\h_{\delta,n}(\omega)&=\frac{1}{n}\log\frac{q-1}{\delta n(n+1)} -\delta \BEF\left(\frac{\omega}{\delta}\right) + \BEF(\omega) - \frac{\log \mathcal A _{\omega n}}{n} \notag \\
\g_n(\delta) &= \inf_{\omega \in (0,\delta]} \h_{\delta,n}(\omega) \notag \quad \text{and} \quad \g^+_n(\delta) = \max\{0,\g_n(\delta)\} \,.
\end{align*}
We also define
\begin{align}\label{eq:h}
\h_{\delta}(\omega)&= -\delta \BEF\left(\frac{\omega}{\delta}\right) + \BEF(\omega) - G(\omega)
\end{align}
so that $\g(\delta)$ defined in \eqref{eq:g} fulfills $\g(\delta)=\inf_{\omega \in (0,\delta]} \h_{\delta}(\epsilon)$.

We start by showing that $\f_{\epsilon,n} \cu \f_{\epsilon}$ on any interval $[a,1]$ such that $0<a<1$. We first show that $\g_n(\delta) \cu \g(\delta)$ on $[a,1]$. To this purpose we write
\begingroup
\allowdisplaybreaks
\begin{align*}
& \sup_{\delta \in [a,1]} | \g_n(\delta) - \g(\delta)| \notag \\ 
&\quad= \sup_{\delta \in [a,1]} \Big| \inf_{\omega \in (0,\delta]} \h_{\delta,n}(\omega) - \inf_{\omega \in (0,\delta]} \h_{\delta}(\omega) \Big| \notag \\
&\quad\stackrel{\mathrm a}{\leq} \sup_{\delta \in [a,1]} \sup_{\omega \in (0,\delta]} | \h_{\delta,n}(\omega) - \h_{\delta}(\omega) | \notag \\
&\quad=  \sup_{\delta \in [a,1]}\sup_{\omega \in (0,\delta]} \Big| -\frac{1}{n}\log \frac{\delta n(n+1)}{q-1} + \frac{\log \mathcal A _{\omega n}}{n} - G(\omega) \Big| \notag \\
&\quad\stackrel{\mathrm b}{\leq} \sup_{\delta \in [a,1]} \Big|\frac{1}{n}\log \frac{\delta n(n+1)}{q-1} \Big| + \sup_{\omega \in (0,\delta]} \Big| \frac{\log \mathcal A _{\omega n}}{n}-G(\omega) \Big| \notag
\end{align*}
\endgroup
where `$\mathrm a$' is due to Lemma~\ref{lemma:absinfbound} and `$\mathrm b$' to triangle inequality. In the last expression, the first addend converges to zero as $n \rightarrow \infty$ since $q$ is constant and $\delta \in [a,1]$ with $a>0$. Moreover, the second addend converges to zero due to the hypothesis that $(1/n) \log \mathcal A _{\omega n} \cu G(\omega)$ and by Lemma~\ref{lemma:supmetric}. Again by Lemma~\ref{lemma:supmetric} we conclude that $\g_n(\delta) \cu \g(\delta)$.

Uniform convergence of $\g_n(\delta)$ to $\g(\delta)$ turns into uniform convergence of $\g^+_n(\delta)$ to $\g^+(\delta)$. In fact, we have $| \g^+_n(\delta) - \g^+(\delta) | \leq | \g_n(\delta) - \g(\delta) |$ for all $\delta$ and $n$, which implies 
\begin{align*}
0\leq \sup_{\delta \in [a,1]} |\g^+_n(\delta) - \g^+(\delta)| \leq \sup_{\delta \in [a,1]} |\g_n(\delta) - \g(\delta)|. 
\end{align*}
By squeeze theorem we have $\sup_{\delta \in [a,1]} |\g^+_n(\delta) - \g^+(\delta)| \rightarrow 0$ as $n \rightarrow \infty$, and therefore $\g^+_n \cu \g^+$ by Lemma~\ref{lemma:supmetric}.

We are now in a position to prove uniform convergence of $\f_{\epsilon,n}$ to $\f_{\epsilon}$. In fact, we have
\begin{align*}
\sup_{\delta \in [a,1]} |\f_{\epsilon,n}(\delta) - \f_{\epsilon}(\delta)| &= \sup_{\delta \in [a,1]} \bigg| -\frac{\log n}{n} + \g^+_n(\delta) - \g^+(\delta) \bigg| \notag \\
&\leq \bigg| \frac{\log n}{n} \bigg| + \sup_{\delta \in [a,1]} | \g^+_n(\delta) - \g^+(\delta) | 
\end{align*}
where we applied triangle inequality. Convergence to zero of the last expression is guaranteed by $\g^+_n \cu \g^+$.

Uniform convergence of $\f_{\epsilon,n}(\delta)$ to $\f_{\epsilon}(\delta)$ leads us to the statement, as follows. Recall that, if $f_n \cu f$ on $A_0$ then $\lim_n \inf_{x \in A_0} f_n(x) = \inf_{x \in A_0} \lim_n f_n(x)=\inf_{x \in A_0} f(x)$, i.e., we can exchange limit and infimum. Hence we can write
\begingroup
\allowdisplaybreaks
\begin{align*}
\lim_{n \rightarrow \infty} - \frac{1}{n} & \log \expectation_{\mathcal{C}(n,r,q)} \left[P_B(\mathsf{C},\epsilon)\right] \notag \\
& \geq \lim_{n \rightarrow \infty} \inf_{\delta \in [a,1]} \f_{\epsilon,n}(\delta) = \inf_{\delta \in [a,1]} \lim_{n \rightarrow \infty} \f_{\epsilon,n}(\delta) \notag \\
&= \inf_{\delta \in [a,1]} \f_{\epsilon}(\delta) \geq \inf_{\delta \in (0,1]} \f_{\epsilon}(\delta) \, .
\end{align*}
\endgroup
In the previous equation array, the first inequality is justified by the fact that if $\alpha_n \rightarrow \alpha$, $\beta_n \rightarrow \beta$, and $\alpha_n \geq \beta_n$ for all $n$ (possibly, larger than some $n_0$), then $\alpha \geq \beta$. Moreover, the two equalities are justified by $\f_{\epsilon,n}(\delta) \cu \f_{\epsilon}(\delta)$. 
\end{IEEEproof}

\begin{remark}\label{eq:null_E}
The function $E_{G}(\epsilon)$ given by \eqref{eq:exp_lower_bound} is nonnegative for all $0<\epsilon<1$, since it is defined as the infimum of the sum of two nonnegative quantities. Moreover, since $E_{G}(\epsilon)$ bounds the error exponent of the given ensemble from below, it must fulfill $E_{G}(\epsilon) = 0$ for all $1-r \leq \epsilon \leq 1$. 
\end{remark}

\begin{remark}
The lower bound $E_G(\epsilon)$ turns out to be useless for all ensembles for which $G(\omega) \rightarrow 0$ as $\omega \rightarrow 0^+$, as for any such ensemble we have $E_G(\epsilon) = 0$ for all $0 < \epsilon < 1$. To see this, simply observe that under this setting we have $\inf_{\omega \in (0,\delta]} \h_{\delta}(\omega) \leq \lim_{\omega \rightarrow 0^+} \h_{\delta}(\omega) = 0$ for all $0 < \delta \leq 1$, and therefore $\g^+(\delta)=0$ for all $0 < \delta \leq 1$. Then, $E_{G}(\epsilon) = \inf_{\delta \in (0,1]} \KL(\delta,\epsilon) =0$ for all $0 < \epsilon < 1$ (simply take $\delta=\epsilon$).
\end{remark}

The following lemma characterizes the function $\g^+(\delta)$ defined in \eqref{eq:gplus}.
\begin{lemma}\label{lemma:gplus}
The function $\g^+(\delta)$ has the following properties: 
\begin{enumerate}
\item $\g^+(\delta)=0$ for all $1-r \leq \delta \leq 1$;
\item If $G(\omega)$ is continuous in $(0,1)$ then $\g^+(\delta)$ is non-increasing and continuous;
\item If $G(\omega)$ is continuous in $(0,1)$ and $\lim_{\omega \rightarrow 0^+} G(\omega) = \gamma < 0$ then:
\begin{enumerate}
\item $\lim_{\delta \rightarrow 0^+} \g^+(\delta) = |\gamma|$;
\item $\delta^* = \sup\{\delta \in (0,1-r] : \g^+(\delta)>0\}$ is strictly positive;
\item $\g^+(\delta)>0$ $\forall \delta \in (0,\delta^*)$; $\g^+(\delta)=0$ $\forall \delta \in[\delta^*,1]$.
\end{enumerate}
\end{enumerate}
\end{lemma}
\begin{IEEEproof}
1) Take any $1-r \leq \delta_1 \leq 1$ and let $\epsilon=\delta_1$. We must have $E_{G}(\delta_1)=\inf_{\delta \in (0,1]} [\KL(\delta,\delta_1)+\g^+(\delta)]=0$ (Remark~\ref{eq:null_E}). This yields $\delta=\delta_1$, hence $\g^+(\delta_1)=E_{G}(\delta_1)=0$.

2) The function $\h(\omega,\delta) = \h_{\delta}(\omega)$ is continuous and derivable \ac{w.r.t.} $\delta$. Since $\partial \h(\omega,\delta)/\partial \delta=\log((\delta-\omega)/\delta)<0$, we have 
\begin{align}\label{eq:h_dec}
\h_{\delta_1}(\omega) > \h_{\delta_2}(\omega) \quad \forall\,\, 0<\omega \leq \delta_1 < \delta_2. 
\end{align}
Moreover, continuity of $G(\omega)$ turns into continuity of $\h_{\delta}(\omega)$ also \ac{w.r.t.} $\omega$. We define $\h_{\delta}(0)=\lim_{\omega \rightarrow 0^+} \h_{\delta}(\omega)$, so that $\h_{\delta}(\omega)$ is continuous (\ac{w.r.t.} $\omega$) on the compact $[0,\delta]$. We let 
$\hat{\omega}_{\delta} = \argmin_{\omega \in [0,\delta]} \h_{\delta}(\omega)$. Taking $z<y$ and using \eqref{eq:h_dec}, we can write $\g(y)=\h_y(\hat{\omega}_y)<\h_y(\hat{\omega}_z)<\h_z(\hat{\omega}_z)=\g(z)$ which shows that $\g(\delta)$ is monotonically decreasing and, as a consequence, that $\g^+(\delta)$ is non-increasing.

Next, we prove continuity of $\g(\delta)$ as it implies continuity of $\g^+(\delta)$. We need to show that for any $\theta>0$ there exists $\alpha(\theta)$ s.t. $|z-y|<\alpha(\theta)$ implies $| \g(z) - \g(y) | < \theta$. It is easy to prove that for any $\theta>0$ there exists $\alpha_1(\theta)$ s.t. $|z-y|<\alpha_1(\theta)$ implies $|\h_z(\omega)-\h_y(\omega)|<\theta/2$ for all $\omega \in (0,\min\{y,z\})$.\footnote{The proof is based on the observation that $|\h_z(\omega)-\h_y(\omega)|$ increases monotonically with $\omega$, yielding $|\h_z(\omega)-\h_y(\omega)| \leq |\h_z(M)-\h_y(M)|=M \BEF(1-|z-y|/M)$, where $M=\max\{z,y\}$. Continuity of $\BEF(\cdot)$ leads to the conclusion.} We refer to this property as ``Property 1''. Moreover, continuity of $\h_{\delta}(\omega)$ w.r.t. $\omega$, ensures that for any $\theta>0$ there exists $\alpha_2(\theta)$ s.t. $|\xi-\omega|<\alpha_2(\theta)$ implies $| \h_{\delta}(\xi) - \h_{\delta}(\omega) | < \theta/2$. We refer to this property as ``Property~2''.

Hereafter we address the case $z<y$, the argument for $z>y$ being very similar. Let $y-z<\min \{\alpha_1(\theta),\alpha_2(\theta)\}$ and recall the above definition of $\hat{\omega}_y$ and $\hat{\omega}_y$. We need to distinguish two cases.

\emph{Case 1:} $\hat{\omega}_y<z$. Property~1 implies $|\h_z(\hat{\omega}_y)-\h_y(\hat{\omega}_y)| < \theta/2$. By \eqref{eq:h_dec} we have $\h_z(\hat{\omega}_y) > \h_y(\hat{\omega}_y)$ and therefore 
\begin{align*}
|\h_z(\hat{\omega}_y)-\h_y(\hat{\omega}_y)| &= \h_z(\hat{\omega}_y)-\h_y(\hat{\omega}_y) \notag \\ 
&= \h_z(\hat{\omega}_y) - \h_z(\hat{\omega}_z) + \h_z(\hat{\omega}_z) - \h_y(\hat{\omega}_y) \notag \\
&\stackrel{\mathrm a}{=} |\h_z(\hat{\omega}_y) - \h_z(\hat{\omega}_z)| + |\h_z(\hat{\omega}_z) - \h_y(\hat{\omega}_y)| \notag
\end{align*}
where `$\mathrm{a}$' is due to the definitions of $\hat{\omega}_z$ and $\hat{\omega}_y$ and to $z<y$. Thus, $|\g(z)-\g(y)| = | \h_z(\hat{\omega}_z) - \h_y(\hat{\omega}_y) | \leq |\h_z(\hat{\omega}_y)-\h_y(\hat{\omega}_y)| < \theta/2 < \theta$.

\emph{Case 2:} $\hat{\omega}_y \geq z$. Property~1 and property~2 imply $|\h_z(z) - \h_y(z)| < \theta/2$ and $|\h_y(z) - \h_y(\hat{\omega}_y)| < \theta/2$, respectively, yielding (by triangle inequality) $|\h_z(z) - \h_y(\hat{\omega}_y)| \leq |\h_z(z) - \h_y(z)| + |\h_y(z) - \h_y(\hat{\omega}_y)| < \theta$. However, we also have 
\begin{align*}
|\h_z(z) - \h_y(\hat{\omega}_y)| &\stackrel{\mathrm a}{=} \h_z(z) - \h_y(\hat{\omega}_y) \notag \\
&= \h_z(z) - \h_z(\hat{\omega}_z) + \h_z(\hat{\omega}_z) - \h_y(\hat{\omega}_y) \notag \\
&\stackrel{\mathrm b}{=} |\h_z(z) - \h_z(\hat{\omega}_z)| + |\h_z(\hat{\omega}_z) - \h_y(\hat{\omega}_y)| \notag 
\end{align*}
where both `$\mathrm a$' and `$\mathrm b$' are due to $\h_z(z) \geq \h_z(\hat{\omega}_z) \geq \h_y(\hat{\omega}_y)$. Hence, $|\g(z) - \g(y)| = |\h_z(\hat{\omega}_z) - \h_y(\hat{\omega}_y)| \leq |\h_z(z) - \h_y(\hat{\omega}_y)| < \theta$.

3a) Let us look at the behavior of $\g^+(\delta)$ as $\delta \rightarrow 0^+$. Since $0 < \omega \leq \delta$, we must also have $\omega \rightarrow 0^+$, which yields $\lim_{\delta \rightarrow 0^+} \g^+(\delta)=\max\{0,\lim_{(\delta,\omega) \rightarrow (0^+,0^+),0<\omega\leq\delta}\h(\omega,\delta)\}=|\gamma|$.

3b) The function $\g^+(\delta)$ tends to a positive number as $\delta \rightarrow 0^+$ and is zero for any $\delta$ between $1-r$ and $1$. Since the function is continuous, $\delta^*=\sup\{\delta \in (0,1-r]: \g^+(\delta)>0\}$ must be strictly positive.

3c) Since $\g^+(\delta)$ is also non-increasing, it must be positive on the whole interval $(0,\delta^*)$ and must be null elsewhere (i.e., on $[\delta^*,1]$).
\end{IEEEproof}

The next theorem shows that, under conditions on $G(\omega)$, there exists an interval of values of $\epsilon$ over which $E_{G}(\epsilon)$ is positive. For the corresponding ensembles, $E_{G}(\epsilon)$ is therefore useful to lower bound $\MLthr$.

\begin{thm}\label{thm:delta_star}
Let $\delta^* = \sup\{\delta \in (0,1-r] : \g^+(\delta)>0\} \leq 1-r$.
If $G(\omega)$ is continuous in $(0,1)$ and $\lim_{\omega\rightarrow 0^+} G(\omega) < 0$, then $E_G(\epsilon) > 0$ $\forall \epsilon \in (0, \delta^*)$ and $E_G(\epsilon) = 0$ $\forall \epsilon \in [\delta^*,1]$, and therefore $\MLthr \geq \delta^*$.
\end{thm}
\begin{IEEEproof}
Take any $\epsilon$ s.t. $\delta^* \leq \epsilon \leq 1$. We have $0 \leq E_{G}(\epsilon)=\inf_{\delta} (\KL(\delta,\epsilon)+\g^+(\delta)) \leq \KL(\epsilon,\epsilon)+\g^+(\epsilon)=0$, and therefore $E_{G}(\epsilon)=0$. Take now any $\epsilon$ s.t. $0 < \epsilon < \delta^*$. Since $\KL(\delta,\epsilon)$ and $\g^+(\delta)$ are both nonnegative functions, to have $E_{G}(\epsilon)=\inf_{\delta} (\KL(\delta,\epsilon)+\g^+(\delta))=0$ we need to find $\delta$ s.t. both $\KL(\delta,\epsilon)$ and $\g^+(\delta)$ are null. To have $\KL(\delta,\epsilon)=0$ we need to choose $\delta=\epsilon$; however, since $0<\epsilon<1-r$ we have $\g^+(\epsilon)>0$ and therefore $E_{G}(\epsilon)>0$.
\end{IEEEproof}

In the next section we present results for two ensembles fulfilling the hypotheses of Theorem~\ref{thm:delta_star}, namely, the ensemble of linear random parity-check codes over $\mathbb F_q$ and the ensemble of fixed-rate binary Raptor codes with linear random precoders \cite{Lazaro16:Raptor}. For the first ensemble the function $E_G(\epsilon)$ can be obtained analytically and coincides with Gallager's random coding bound over the \ac{$q$-EC}. For the second one, $E_G(\epsilon)$ shall be computed numerically. However, if only the lower bound on $\MLthr$ is of interest, it may be computed  by simply solving a $2 \times 2$ system of equations.

\section{Results for Specific Ensembles}\label{sec:examples}
\subsection{Linear Random Parity-Check Codes}

Consider the ensemble of linear random parity-check codes over $\field_q$ induced by an $(1-r)n \times n$ random parity-check matrix whose entries are \ac{i.i.d.} random variables uniformly distributed in $\field_q$. For this ensemble we have the following result.

\begin{thm}
For the ensemble of linear random parity-check codes we have $\delta^*=1-r$ and therefore $\MLthr=1-r$. Moreover
\begin{equation}\label{eq:rce}
E_{G}(\epsilon)=\left\{ \begin{array}{ll}
-\log \big(\frac{1-\epsilon}{q}+ \epsilon\big)-r\log q & \,\,0< \epsilon< \epsilon_c\\
\mathcal{D}(1-r,\epsilon) & \,\,\epsilon_c \leq \epsilon < 1-r\\
0 & \,\,\epsilon \geq 1-r
\end{array}\right. 
\end{equation}
where $\epsilon_c = (1-r)/(1+(q-1)r)$.
\end{thm}
\begin{IEEEproof}
The expected weight enumerator of the linear random parity-check ensemble is $\mathcal A_{\omega n} = {n \choose \omega n} (q-1)^{\omega n} q^{-(1-r)n}$ and the corresponding weight spectral shape is $G(\omega) = \BEF(\omega) +\omega \log(q-1) - (1-r) \log q$. Uniform convergence of $\frac{1}{n} \log \mathcal A_{\omega n}$ to $G(\omega)$ may be proved in a very simple way, by observing that
\begin{align*}
\sup_{\omega} \Big| \frac{1}{n} \log \mathcal A_{\omega n} - G(\omega) \Big| &= \sup_{\omega} \Big| \frac{1}{n} \log {n \choose \omega n} - \BEF(\omega) \Big| \notag \\
                                                                                                                          &\leq \sup_{\omega} \Big| \frac{\log(n+1)}{n} \Big| = \Big| \frac{\log(n+1)}{n} \Big| \notag
\end{align*}
where we applied the lower bound in \eqref{eq:bin_coeff_bounds}. Since $| \frac{\log(n+1)}{n} | \rightarrow 0$ as $n \rightarrow \infty$, we conclude that $\frac{1}{n} \log \mathcal A_{\omega n} \cu G(\omega)$.

The function $\h_{\delta}(\omega)$ defined in \eqref{eq:h} assumes the form
\begin{align*}
\h_{\delta}(\omega) = -\delta \BEF\left(\frac{\omega}{\delta}\right) - \omega \log(q-1) + (1-r) \log q \, .
\end{align*}
Let $\hat{\omega}(\delta)=\frac{q-1}{q}\delta$. It is easy to see that this function tends to $(1-r)\log q$ when $\omega \rightarrow 0^+$, is monotonically decreasing for $\omega \in (0,\hat{\omega}(\delta))$, takes a minimum at $\omega=\hat{\omega}(\delta)$, and increases monotonically for $\omega \in (\hat{\omega}(\delta),\delta]$. Hence, we have $\g(\omega) = \inf_{\omega\in(0,\delta]} \h_\delta(\omega) = \h_{\delta}(\hat{\omega}(\delta)) = (1-r-\delta)\log q$ so that
\begin{align*}
\g^+(\delta)\! = \max \{0,\g(\delta) \} =\! \left\{ \begin{array}{cl} \!(1-r-\delta)\log q & \mathrm{if}\, 0<\delta<1-r \\ \!0 & \mathrm{if}\, 1-r\leq\delta<1. \end{array} \right.
\end{align*}
The parameter $\delta^*$ is therefore equal to $1-r$. Since $\MLthr \geq \delta^*=1-r$ and $\MLthr \leq 1-r$, we obtain $\MLthr=1-r$.

Next, we develop $E_{G}(\epsilon)$ analytically. Based on the above findings, we have
\begin{align}\label{eq:E_linear_random}
E_{G}(\epsilon) 
                         &= \min \Big\{\inf_{\delta \in (0,1-r)} [\KL(\delta,\epsilon) + (1-r-\delta)\log q ], \notag \\
                         &\qquad\qquad \inf_{\delta \in [1-r,1] } \KL(\delta,\epsilon)\Big\}
\end{align}
that immediately yields $E_{G}(\epsilon)=0$ for all $\epsilon\geq 1-r$ (it suffices to take $\delta=\epsilon$), corresponding to the third row of \eqref{eq:rce}. For $0<\epsilon <1-r$ we need to analyze the function $\f_{\epsilon}(\delta)=\KL(\delta,\epsilon)+\g^+(\delta)=\KL(\delta,\epsilon) + (1-r-\delta)\log q$. Let $\hat{\delta}(\epsilon)=\frac{q \epsilon}{1+(q-1)\epsilon}$. Taking the derivative with respect to $\delta$, it is immediate to see that this function decreases monotonically for $\delta<\hat{\delta}(\epsilon)$, takes a minimum at $\delta=\hat{\delta}(\epsilon)$, and increases monotonically for $\delta>\hat{\delta}(\epsilon)$. Hereafter we need to distinguish the two cases $\hat{\delta}(\epsilon)<1-r$ and $\hat{\delta}(\epsilon)\geq1-r$. It is immediate to verify that they correspond to $0 < \epsilon < \epsilon_c$ and $\epsilon_c \leq \epsilon < 1-r$, respectively, where $\epsilon_c = (1-r)/(1+(q-1)r)$.

\emph{Case 1: $0 < \epsilon < \epsilon_c$}. In this case the function $\KL(\delta,\epsilon) + (1-r-\delta)\log q$ has a minimum at $\delta=\hat{\delta}(\epsilon)$. It takes the value $\KL(1-r,\epsilon)$ at $\delta=1-r$. Therefore we obtain
\begin{align*}
E_{G}(\epsilon) &= \min \{\KL(\hat{\delta}(\epsilon),\epsilon) + (1-r-\hat{\delta}(\epsilon))\log q,\KL(1-r,\epsilon) \} \notag \\
                         &= \KL(\hat{\delta}(\epsilon),\epsilon) + (1-r-\hat{\delta}(\epsilon))\log q \notag \\
                         &= -\log\Big(\frac{1-\epsilon}{q}+\epsilon \Big) - r \log q
\end{align*}
where the third expression follows from simple algebraic manipulation. This yields the first row of \eqref{eq:rce}.

\emph{Case 2: $\epsilon_c < \epsilon < 1-r$}. In this case the function $\KL(\delta,\epsilon) + (1-r-\delta)\log q$ is monotonically decreasing for $\delta \in (0,1-r)$, so its infimum is taken as $\delta \rightarrow (1-r)^{-}$. We obtain $E_{G}(\epsilon)=\min\{\KL(1-r,\epsilon),\KL(1-r,\epsilon)\} =\KL(1-r,\epsilon)$ that corresponds to the second row of \eqref{eq:rce}.
\end{IEEEproof}
\begin{remark} Interestingly, the expression \eqref{eq:rce} of $E_{G}(\epsilon)$ turns out to coincide with that of Gallager's random coding error exponent for the \ac{$q$-EC} \cite{qEras:ISIT2008}.
\end{remark}

\subsection{Fixed-Rate Raptor Codes with Linear Random Precoders}

In this subsection we consider binary fixed-rate Raptor code ensembles with linear random precoding. A vector of $rn$ information bits is first encoded by an outer linear block code picked randomly in the ensemble of binary linear random parity-check codes with design rate $\ro$, providing a vector of $rn/\ro$ intermediate bits. Intermediate bits are further encoded by an inner fixed-rate \ac{LT} code of rate $\ri$ and output degree distribution $\Omega(x)=\sum_j \Omega_j x^j$, generating $n$ encoded bits. The overall design rate is $r=\ro \ri$.

The weight spectral shape of this ensemble was characterized in \cite{Lazaro16:Raptor}. It is given by
\begin{align}\label{eq:G_raptor}
G(\omega) = \BEF(\omega) - \ri(1-\ro) - \nu_{\omega}(\lambda_0)
\end{align}
where 
\begin{align}\label{eq:fmax}
\!\!\!\!\!\!\!\!\nu_{\omega}(\lambda) =  \BEF(\lambda) + \omega \log(\rho(\lambda)) + (1-\omega)\log(1-\rho(\lambda))
\end{align}
and
\begin{align}\label{eq:lambda_0}
\lambda_0 = \lambda_0(\omega) = \argmax_{\lambda \in \mathcal{D}} \nu_{\omega}(\lambda).
\end{align}
In \eqref{eq:lambda_0}, $\mathcal{D}=[0,1)$ if $\Omega_j=0$ for any even $j$ and $\mathcal{D}=(0,1)$ otherwise. Moreover, $\rho(\lambda)=\frac{1}{2} \sum_{j=1}^{d} \Omega_j [1-(1-2\lambda)^j]$, being $d$ the maximum \ac{LT} output degree. Again from \cite{Lazaro16:Raptor}:
\begin{enumerate}
\item $G(\omega)$ in \eqref{eq:G_raptor} is continuous.
\item $\lim_{\omega \rightarrow 0^+} G(\omega)<0$ iff $(\ri,\ro) \in \cP$, where 
\begin{align}\label{eq:P}
\cP=\Big\{&(\ri,\ro) \succeq (0,0): \ri(1-\ro) \notag \\ 
& > \max_{\lambda\in\mathcal{D}} [ \ri \BEF(\lambda) +  \log(1-\rho(\lambda))] \Big\}.
\end{align}
\item The derivative of $G(\omega)$ is
\begin{align}\label{eq:G_der_raptor}
G'(\omega) = \log\frac{1-\omega}{\omega} + \log\frac{\rho(\lambda_0)}{1-\rho(\lambda_0)}\,.
\end{align}
\item $G'(\omega)>0$ for $0<\omega<\frac{1}{2}$ and $\lim_{\omega \rightarrow 0^+}G'(\omega)=+\infty$.
\end{enumerate}

Uniform convergence of $\frac{1}{n} \log \mathcal A_{\omega n}$ to $G(\omega)$ can be proved using arguments from \cite[Sec.~III]{Lazaro16:Raptor}. Moreover, the hypotheses of Theorem~\ref{thm:delta_star} are satisfied when $(\ri,\ro)\in\cP$, where $\cP$ is given by \eqref{eq:P}. As opposed to linear random parity-check ensembles, in this case $E_{G}(\epsilon)$ shall be computed numerically. However, if only the lower bound $\delta^*$ on the \ac{ML} decoding threshold $\MLthr$ is of interest, it may be computed efficiently, as shown next.

\begin{thm}
Consider a binary Raptor ensemble with a linear random precoder and let $(\ri,\ro) \in \cP$. Then $\MLthr \geq \delta^*$ where $\delta^*$ is the smallest $\hat{\delta}$ s.t. $(\hat{\delta},\hat{\lambda}_0)$ is a solution of the $2 \times 2$ system
\begin{align}
\!\!\!\!\!\ri(1-\ro) -\ri \BEF(\hat{\lambda}_0) - (1-\hat{\delta}) \log(1-\rho(\hat{\lambda}_0)) &= 0 \label{eq:raptor_thm_1} \\
\ri \log\frac{1-\hat{\lambda}_0}{\hat{\lambda}_0} - \frac{1-\hat{\delta}}{1-\rho(\hat{\lambda}_0)} \rho'(\hat{\lambda}_0)  \log e &= 0 \, . \label{eq:raptor_thm_2}
\end{align}
\end{thm}
\begin{IEEEproof}
If $(\ri,\ro) \in \cP$ then Theorem~\ref{thm:delta_star} applies. We have $\MLthr \geq \delta^*$, where $\delta^*=\sup\{\delta \in (0,1-r]: \g^+(\delta)>0\}$ and $\delta^*>0$. Owing to continuity of $\h_{\delta}(\omega)$ we can write $\g^+(\delta)=\max\{0,\h_{\delta}(\hat{\omega})\}$, where $\hat{\omega}=\hat{\omega}(\delta)=\argmax_{\omega \in (0,\delta]} \h_{\delta}(\omega)$. From \eqref{eq:h} and \eqref{eq:G_der_raptor} we obtain
\begin{align}\label{eq:dh_raptor}
\frac{\mathrm d \h_{\delta}(\omega)}{\mathrm d \omega} &= \log\frac{\omega}{\delta-\omega} -\log\frac{\rho(\lambda_0)}{1-\rho(\lambda_0)}
\end{align}
which reveals how $\mathrm d \h_{\delta}(\omega) / \mathrm d \omega \rightarrow +\infty$ as $\omega \rightarrow \delta^-$.\footnote{$\rho(\lambda_0(\omega))$ cannot converge to $1$ as $\omega \rightarrow \delta^-$ for any $0 < \delta \leq 1-r$.} Thus, the maximum cannot be taken at $\omega=\delta$ and $\hat{\omega}$ must be a solution of $\mathrm d \h_{\delta}(\omega) / \mathrm d \omega = 0$. Defining $\hat{\lambda}_0=\argmax_{\lambda\in\mathcal{D}} \nu_{\hat{\omega}}(\lambda)$ and recalling \eqref{eq:dh_raptor}, after some algebraic manipulation this translates to 
\begin{align}\label{eq:homega_hlambda}
\hat{\omega} = \delta \rho(\hat{\lambda}_0) \, .
\end{align}
The parameter $\hat{\lambda}_0$ must be a solution of $\mathrm d \nu_{\hat{\omega}}(\lambda) / \mathrm d \lambda = 0$. Developing the derivative we obtain
\begin{align}\label{eq:dnu}
\ri \log\frac{1-\hat{\lambda}_0}{\hat{\lambda}_0} + \hat{\omega} \frac{\rho'(\hat{\lambda}_0)}{\rho(\hat{\lambda}_0)} \log e - (1-\hat{\omega}) \frac{\rho'(\hat{\lambda}_0)}{1-\rho(\hat{\lambda}_0)} \log e = 0 .
\end{align}
So far we have shown that $\g^+(\delta)=\max\{0,\h_{\delta}(\hat{\omega})\}$ where $(\hat{\omega},\hat{\lambda}_0)$ is a solution of the system of simultaneous equations \eqref{eq:homega_hlambda} and \eqref{eq:dnu}. Recall now from Theorem~\ref{thm:delta_star} that $\g^+(\delta)>0$ for all $0<\delta<\delta^*$ and $\g^+(\delta)=0$ for all $\delta^* \leq \delta \leq 1$. This necessarily implies $\h_{\delta}(\hat{\omega})>0$ for all $0 < \delta < \delta^*$ and $\h_{\delta^*}(\hat{\omega})=0$, i.e., $\delta^*$ is the smallest $\hat{\delta}$ such that $\h_{\hat{\delta}}(\hat{\omega})=0$, i.e., after simple manipulation, the smallest $\hat{\delta}$ such that
\begin{align*}
\ri(1-\ro) - \ri \BEF(\hat{\lambda}_0) - (1-\hat{\delta}) - (1-\hat{\delta}) \log \Big(1-\frac{\hat{\omega}}{\hat{\delta}} \Big) = 0 .
\end{align*} 
Substituting \eqref{eq:homega_hlambda} (with $\delta=\hat{\delta}$) into this latter equation yields \eqref{eq:raptor_thm_1}, while substituting it into \eqref{eq:dnu} yields \eqref{eq:raptor_thm_2}.
\end{IEEEproof}
\begin{example}
Let $\ro=0.99$, $\ri=0.8$, and $\Omega(x)$ be the \ac{LT} output distribution of 3GPP Raptor codes, i.e.,
\begin{align*}
\Omega(x) &= 0.0098 x + 0.4590 x^2 + 0.2110 x^3 + 0.1134 x^4 \notag \\ 
&+ 0.1113 x^{10} + 0.0799 x^{11} + 0.0156 x^{40} \, .
\end{align*}
By direct calculation one can verify that $(\ri,\ro) \in \cP$. Solving \eqref{eq:raptor_thm_1} and \eqref{eq:raptor_thm_2} we obtain the unique solution $(\hat{\delta},\hat{\lambda}_0)=(0.090771,0.009951)$ from which we conclude that $\MLthr \geq \delta^* = 0.090771$. This bound is relatively tighter that the one obtained by employing the general bound in \cite{Shulman99:random}, which returns $\MLthr \geq 0.003827$.
\end{example}

\newpage
\section{Conclusions}\label{sec:conc}

A lower bound on the \ac{ML} error exponent of linear code ensembles over erasure channels has been derived.
The lower bound requires, under mild conditions, just the knowledge of the ensemble weight spectral shape. The application to some linear block code ensembles has been demonstrated. For the specific case of fixed-rate Raptor code ensembles, the bound allows to compute a lower bound on the \ac{ML} decoding threshold, that is remarkably tighter with respect to the lower bound obtained  with established techniques.


\appendices



\bibliography{IEEEabrv,qEC}

\end{document}